\documentclass[prb,twocolumn,showpacs,preprintnumbers,amsmath,amssymb]{revtex4}

\usepackage{graphicx}
\usepackage{dcolumn}
\usepackage{epstopdf}
\begin{document}
\title{Extended paraconductivity regime in underdoped cuprates}

\author{S. Caprara$^1$, M. Grilli$^1$, B. Leridon$^2$, and J. Lesueur$^2$}

\affiliation{$^1$Istituto Nazionale per la Fisica della Materia, 
Unit\`a Roma 1 and SMC Center,\\ and Dipartimento di Fisica
Universit\`a di Roma ``La Sapienza'', piazzale Aldo Moro 5, I-00185 Roma, 
Italy\\
$^2$Laboratoire de Physique Quantique - ESPCI/UPR5-CNRS,
10, Rue Vauquelin - 75005 Paris - France}



\begin{abstract}
We reconsider transport experiments in strongly anisotropic superconducting
cuprates and we find that universal Aslamazov-Larkin (AL)
paraconductivity in two dimensions
is surprisingly  robust even in the underdoped regime
below the pseudogap crossover temperature $T^*$. We also establish that the
underlying normal state resistivity in the pseudogap phase is (almost) linear in 
temperature, with all the deviations being quantitatively accounted by
AL paraconductivity. The disappearence of paraconductivity is governed by the
disappearence of gaussian pair fluctuations at an energy scale related to $T^*$. 
\end{abstract}

\date{\today}
\maketitle
  
Recent transport experiments\cite{Luo,leridon1,leridon2} in superconducting   
cuprates have shown that the paraconductivity effects in the normal state 
close to the critical temperature $T_c$ are well described by the following 
expressions valid in two and three dimensions, respectively,  
\begin{eqnarray}  
\Delta \sigma_{D=2}^{exp}&=& \frac{e^2}{16\hbar d\varepsilon_0  
\sinh( \varepsilon /\varepsilon_0)}, \label{sigmaEXP2D} \\  
\Delta \sigma_{D=3}^{exp}&=& \frac{e^2}{16 \hbar \xi_{c0}
{\sqrt{2\varepsilon_0\sinh( 2\varepsilon/ \varepsilon_0)}}},
\label{sigmaEXP3D}  
\end{eqnarray} 
where $d$ is the distance between ${\rm CuO_2}$ layers, $\xi_{c0}$ is the 
coherence length along the direction perpendicular to the layers, 
$\varepsilon \equiv \log(T/T_c)$, and $\varepsilon_0 \equiv \log(T^\#/T_c)$. 
Here $T^\#$ is a temperature scale which increases with decreasing doping and 
appears to follow the characteristic crossover temperature  $T^*$ below which 
many different experiments in the cuprates detect a pseudogap 
opening\cite{TIMUSK}. The above expressions (and the experimental data that 
they fit well) display two remarkable features. First of all, close to $T_c$, 
for small values of $\varepsilon$, they reproduce the Aslamazov-Larkin (AL)  
form of paraconductivity\cite{AL}
\begin{eqnarray}  
\Delta \sigma^{AL}_{D=2}&=&   
\frac{e^2}{16\hbar d\varepsilon}, \label{sigmaAL2D}\\  
\Delta \sigma^{AL}_{D=3}&=&   
\frac{e^2}{32 \hbar \xi_{c0}\hbar \sqrt{\varepsilon}}.  
 \label{sigmaAL3D}  
\end{eqnarray}  
These expressions account well for the fluctuating regime near $T_c$ both in   
optimally and underdoped cuprates, with ${\rm YBaCuO_{6+x}}$ (YBCO) 
displaying three-dimensional ($3D$) fluctuations, whereas the other more 
anisotropic compounds (LSCO and BSCCO) have a two-dimensional ($2D$) behavior. 
The fact that the paraconductivity in strongly anisotropic (quasi-$2D$) 
underdoped cuprates is described by ``traditional'' AL fluctuations is at odds 
with the widespread idea that below the pseudogap formation temperature $T^*$ 
particle-particle pairs are formed, which only become phase-coherent at the 
lower superconducting transition temperature $T_c$. According to this picture, 
below the temperature of pair formation the fluctuations would be 
vortex-driven and should display a Kosterliz-Thouless behavior, with 
exponential temperature dependences. On the contrary, it seems a 
well-established experimental fact that the superconducting fluctuations in 
the more $2D$-like systems (essentially all, but the YBCO) display AL 
power-law behaviors in 
$\varepsilon$\cite{Balestrino,Balestrino92,Lang,cimberle,Varlamov,Curras}. 
Remarkably, in 
$D=2$ the AL theory of paraconductivity does not allow for any fitting 
parameter besides the experimentally well accessible distance between the $2D$ 
layers, which translates the $2D$ conductivity, with dimensions $\Omega^{-1}$, 
in a $3D$ conductivity with dimensions $\Omega^{-1}$m$^{-1}$. Therefore the AL 
paraconductive behavior observed near $T_c$ strikingly shows that the 
establishment of superconducting phase coherence in these materials 
{\it is not due to a simple condensation of preformed pairs}. This by no means 
implies that preformed pairs are not present below $T^*$, but simply means 
that the superconducting coherence is driven by the formation of more loosely 
bound traditional BCS pairs. Various proposals have already been put forward 
based on the coexistence of fermionic quasiparticles (eventually forming BCS 
pairs at $T_c$) and more or less bosonic preformed 
pairs\cite{geshkenbein,twogap,ranninger,piegari}.   
  
The second remarkable features of the experiments described by Eqs.
(\ref{sigmaEXP2D}) and (\ref{sigmaEXP3D}) regards the exponential suppression 
of the paraconductivity  when $\varepsilon>\varepsilon_0$. While it is quite 
natural that superconducting fluctuations decay when moving away from $T_c$, 
no longer contributing to the conductivity, the fact that AL fluctuations 
survive up to $T^*$ is surprising. In underdoped cuprates this rapid drop in 
the AL fluctuations occurs at the temperature scale $T^\#\sim T^*$, which is 
substantially higher than the superconducting temperature $T_c$. 

In principle one could argue that $T^*$ is indeed the temperature below which 
superconducting Cooper-pair fluctuations arise, and therefore it is not 
surprising that they contribute {\it \'a la} AL to the paraconductivity. 
However, upon underdoping, $T^*$ increases, while $T_c$ decreases. If this is 
interpreted within a standard scheme of strong-coupling pairing, the phase 
fluctuations would be (the only) responsible for paraconductivity, in $D=2$, 
and one should rather observe the Kosterliz-Thouless-like condensation of 
preformed pairs. 

In this paper we focus on these two main features of the paraconductivity
experiments. Firstly we critically reexamine the AL theory and the possible 
occurrence of momentum and/or energy cutoffs in the critical pair fluctuations.
This will provide a different perspective on the rapid drop of the
paraconductivity above $T^*$, with respect to previous 
works\cite{vidal,Mishonov}, but will leave open the question of the mechanism 
allowing for the long survival of AL fluctuations in the pseudogap phase of 
underdoped cuprates. Then we will focus on the 2D materials and examine the 
robustness of the AL paraconductivity at various doping upon varying the
assumed normal-state resistivity. Again, our scope is neither to provide a
microscopic theory for the normal-state phase nor for its interplay with pair 
fluctuations below $T^*$. Our main concern here is to extract the most likely 
form of the normal-state resistivity in connection to the distinct presence of 
paraconductivity. The 2D case is the only one of our concern because the 
universal form of AL paraconductivity renders this analysis more stringent.

{\it --- Paraconductivity suppression around $T^*$ ---}
We discuss the paraconductivity starting by revisiting the derivation of the 
standard AL result in $D$ dimensions\cite{AL,larkin-varlamov}
\begin{eqnarray}  
\Delta\sigma_D^{AL} &=& \alpha_D\int\frac{d^D{\mathbf q}}{(2\pi)^D}~q^2  
~{\mathcal I}(\Omega_{\mathbf q};T),\label{basic}\\  
{\mathcal I}(\Omega_{\mathbf q};T)&\equiv&  
\int_{-\infty}^{+\infty}\frac{dz}{\pi }  
\frac{z^2}{(z^2+\Omega_{\mathbf q}^2)^2}\left[  
-\frac{\partial b(z)}{\partial z}\right]\label{basica},  
\end{eqnarray}  
where $\alpha_D$ is a prefactor which acts as a coupling constant of the 
collective pair fluctuations with the electromagnetic field and is related to 
the fermionic loops in the diagrammatic approach\cite{AL,larkin-varlamov} 
calculated at zero external frequency. $\Omega_{\mathbf q}$ is the inverse 
relaxation time of the collective pair fluctuations with a wavevector 
${\mathbf q}$, which at low momenta takes the hydrodynamic form 
$\Omega_{\mathbf q}\approx m+\nu q^2$ with a ``mass'' term 
$m\propto T\log(T/T_c)\propto T-T_c$ measuring the distance from criticality, 
and a characteristic inverse time scale $\nu$; $b(z)=[e^{z/T}-1]^{-1}$ is the 
Bose distribution at a temperature $T$ (in energy units). Here and in the 
following $q\equiv |{\mathbf q}|$, we take $\hbar=1$, and measure lengths and 
inverse wavevectors in units of the lattice spacing $a$. The inverse 
relaxation time $\Omega_{\mathbf q}$ is often referred to as the energy of the 
collective pair fluctuations. Although this terminology is improper, as the 
dynamics of pair fluctuations is relaxational and not propagating, we adopt it 
hereafter for the sake of definiteness. To make contact with Ref. 
\onlinecite{AL} the prefactor within the AL theory is 
$\alpha_D=16 e^2 \nu^2/D$, the mass term is $m=\gamma^{-1}\log(T/T_c)$, and 
the characteristic inverse time scale is $\nu\simeq\gamma^{-1}\xi_0^2$, where 
$\gamma=\pi/(8T)\simeq \pi/(8 T_c)$ is a characteristic time scale for the   
damping of pair fluctuations, and $\xi_0$ is the coherence length (in units of
the lattice spacing). We point out that the above Eqs. (\ref{basic}) and 
(\ref{basica}) are valid within a Ginzburg-Landau (GL) context, under quite 
general conditions, for a generic expression of $\Omega_{\mathbf q}$, which 
may include corrections to the hydrodynamic expression at higher momenta. For 
instance, in a lattice system,  both the factor $q^2$ and the expression for 
$\Omega_{\mathbf q}$ are replaced by suitable generalizations which preserve 
the lattice periodicity.  
  
The suppression of paraconductivity could in principle arise from, e.g., the 
subleading temperature dependence of the prefactors $\alpha_D$ and from the
subleading temperature dependence of the integral in Eq. (\ref{basica}). This 
analysis was carried out previously\cite{reggiani,NOUS}, finding power-law 
dependencies in $\varepsilon$. However, the suppression of the paraconductvity 
at higher temperature is by far sharper than the one provided by the 
temperature as the natural cutoff. We are therefore led to discuss the role of 
an intrinsic cutoff for the momentum integral in Eq. (\ref{basic}). The 
analytical development within a BCS derivation of the effective GL theory 
leads to a natural momentum cutoff $\sim \xi_0^{-1}$ for higher momenta. This 
cutoff can alternatively be described as the appearance of higher-order terms 
in the $q$ dependence of $\Omega_{\mathbf q}$, beyond the lowest-order term 
$\sim q^2$. However, neither a strict momentum cutoff $q\le q_C\sim\xi_0^{-1}$,
nor the introduction, e.g., of a $q^4$ term in $\Omega_{\mathbf q}$ account 
for the observed behavior of the paraconductivity\cite{silva}.

Based on physical arguments, it was proposed\cite{vidal,Mishonov} that, 
rather than a strict cutoff on $q$, a cutoff should be imposed on the energy 
(namely, the inverse relaxation time) of the collective pair fluctuations. 
This cutoff, within the standard BCS-GL theory, takes the form 
$m+\nu q^2\le \xi_0^{-2}T_c$ and leads to a sharper reduction with respect to 
a strict momentum cutoff. This is easily understood by considering that, away 
from $T_c$, $m$ increases, so that a strict cutoff $\Omega_C$ on   
$\Omega_{\mathbf q}\approx m+\nu q^2$ amounts to a strict momentum cutoff   
$q^2\le q_C^2\equiv(\Omega_C-m)/\nu$, which decreases with increasing 
temperature, thus shrinking the region of momenta which contribute to the 
paraconductivity. This effect adds on top of the reduction associated with an 
increasing mass $m$, and determines a more rapid decrease at higher 
temperatures. Nevertheless the experimental suppression of the 
paraconductivity, fitted by Eqs. (\ref{sigmaEXP2D}) and (\ref{sigmaEXP3D}),
is even stronger. Therefore in the remaining part of this section, we analyze 
the experimental data within a framework which, although related to the 
presence of an energy cutoff, rather relies on a model for an effective 
``density of states'' of the collective pair fluctuations. Indeed, we 
transform the momentum integral into an energy integral, by introducing the   
effective density of states 
\[  
{\mathcal N}_D(\Omega)=\int\frac{d^D{\mathbf q}}{(2\pi)^D}~q^2  
~\delta(\Omega-\Omega_{\mathbf q})  
\]  
for an arbitrary expression of $\Omega_{\mathbf q}$ as a function of the   
momentum. This includes as particular cases, e.g., the effect of a   
higher-order momentum dependence of $\Omega_{\mathbf q}$ with respect to the   
hydrodynamic $q^2$ dependence, and/or the cutoff condition   
$\Omega_{\mathbf q}\le \Omega_C$. We observe that the minimum value for   
$\Omega_{\mathbf q}$ is $m$, and therefore  
\begin{equation}  
\Delta\sigma_D= \alpha_D\int_{m}^{+\infty}d\Omega~   
{\mathcal N}_D(\Omega) ~{\mathcal I}(\Omega;T).  
\label{starting}  
\end{equation}  
This equation is our starting point. For the sake of simplicity we
discuss the case in which ${\mathcal I}(\Omega_{\mathbf q};T)$ has the leading 
AL expression ${\mathcal I}(\Omega_{\mathbf q};T)=T/(2\Omega_{\mathbf q}^3)$, 
but the analysis can be easily extended to the case in which 
${\mathcal I}(\Omega;T)$ assumes a more complicated dependence on 
$\Omega_{\mathbf q}$ which interpolates between the low-$T$ 
$(\Omega_{\mathbf q}\gg T)$ and the high-$T$ $(\Omega_{\mathbf q}\ll T)$ 
regimes\cite{NOUS}.

A sharp energy cutoff $\Omega\le\Omega_C$ translates into a vanishing   
DOS, ${\mathcal N}_D(\Omega)\equiv 0$ for $\Omega>\Omega_C$. We relax this   
condition, and only assume that the DOS vanishes at infinity. More precisely,  
we write the function to be integrated in Eq. (\ref{starting}) as the  
derivative of an auxiliary function,  
$\alpha_D{\mathcal N}_D(\Omega) {\mathcal I}(\Omega;T)\equiv   
-{\mathcal F}_D'(\Omega)$, with $T$ taken as a parameter, and assume that  
${\mathcal F}_D(\Omega)$ vanishes as $\Omega\to +\infty$. Then, evidently  
$\Delta\sigma_D={\mathcal F}_D(m)$. Recalling that 
$m=\gamma^{-1}\varepsilon$, with $\varepsilon\equiv \log(T/T_c)$, we can 
extract ${\mathcal N}_D(\Omega)$ from the interpolating formula for the 
paraconductivity proposed in Refs. \onlinecite{leridon1,leridon2}, Eq. 
(\ref{sigmaEXP2D}) for $D=2$ and Eq. (\ref{sigmaEXP3D}) for $D=3$.
Thus we find  
$$
{\mathcal N}_D(\Omega)=-\frac{1}{\alpha_D {\mathcal I}(\Omega;T)}  
\frac{d}{d \Omega}\Delta\sigma_D^{exp}(\varepsilon=\gamma\Omega).
$$
 
Therefore, we are led to the conclusion that the spectrum of the inverse 
relaxation time for the collective pair fluctuations is cut off exponentially 
at higher $\Omega$ and the characteristic scale for this suppression, 
$\Omega_0\equiv \gamma^{-1}\varepsilon_0$, increases with decreasing doping, 
following $T^*$. The presence of this scale is highly significant and rises 
the issue of the relation between Cooper pair fluctuations and 
pseudogap\cite{notavidal}. 
Since the microscopic interpretation of this finding is 
beyond the scope of the present work, here we only illustrate two possible 
interpretations. Coming from high temperatures $T>T^*$, one can idenjtify 
$T^*$ as the mean-field-like temperature for superconductivity, below which 
the fluctuations bring the critical temperature down to $T_c$. The bifurcation 
between $T^*$ and $T_c$ around optimal doping can be interpreted in a gaussian 
GL scheme within a two-gap model\cite{twogap}. 

An alternative interpretation can be proposed starting from $T_c$ as the
temperature above which pair fluctuations set in. The disappearence of pair 
fluctuations above $T^*$ can here be interpreted as due to some additional 
mechanism of strong mixing between the particle-.particle and the 
particle-hole channels. In particular, within a scenario with a quantum 
critical point around optimal doping, the region above $T^*$ is characterized 
by the presence of quantum-critical fluctuations, which can couple to the 
superconducting fluctuations, and suppress them. These two possibilities are 
presently under investigation\cite{NOUS}.
   
{\it --- Aslamazov-Larkin paraconductivity in the pseudogap phase ---}
The occurrence of the AL paraconductivity  is particularly stringent in 2D 
systems, where the AL paraconductivity does not contain fitting parameters and 
takes a universal form with a power-law dependence in $\varepsilon$ and a 
definite prefactor [see Eq. (\ref{sigmaAL2D})]. For this reason, here we 
concentrate on 2D BSCCO compounds. The choice of a normal-state conductivity 
(or resistivity) becomes rather natural around optimal doping, where 
$\rho_n(T)$ is linear over a wide temperature range. It is in this case that 
the presence of a AL paraconductivity becomes particularly clear both in 
$D=2$\cite{Balestrino,Lang,cimberle,Varlamov,Curras} and 
$D=3$\cite{leridon1,Freitas,carballeira}. Remarkably, since the 
paraconductivity 
diverges at $T_c$, the choice of a specific (finite) normal state conductivity 
$\sigma_n(T)$ affects little the total conductivity
$\sigma(T)=\sigma_n(T)+\Delta\sigma^{AL}(T)$,
and the divergence of $\Delta\sigma^{AL}$ can not be missed by a wrong choice 
of the 
normal state. However, the choice of the correct $\sigma_n$ becomes crucial 
for the correct description of the paraconductivity away from $T_c$. Therefore 
we here systematically investigate how different normal-state resistivities 
affect the determination of $\Delta\sigma^{AL}$ in the resistivity data of 
Ref. \onlinecite{WATANABE}. First of all, we notice (see Fig. 2 in Ref.
\onlinecite{WATANABE}) that above a temperature $T^*$ the resistivity is 
linear in temperature, while it acquires a downward curvature at lower 
temperatures. Therefore we assume the normal state resistivity $\rho_n$ to be 
described by a straight line above $T^*$, while a quadratic curve is taken 
below it. To explore the effects of assuming different  $\rho_n$ we take below 
$T^*$ the set of parabol\ae  reported in Fig. 1(a).
\begin{figure}  
\includegraphics[angle=-90,scale=0.4]{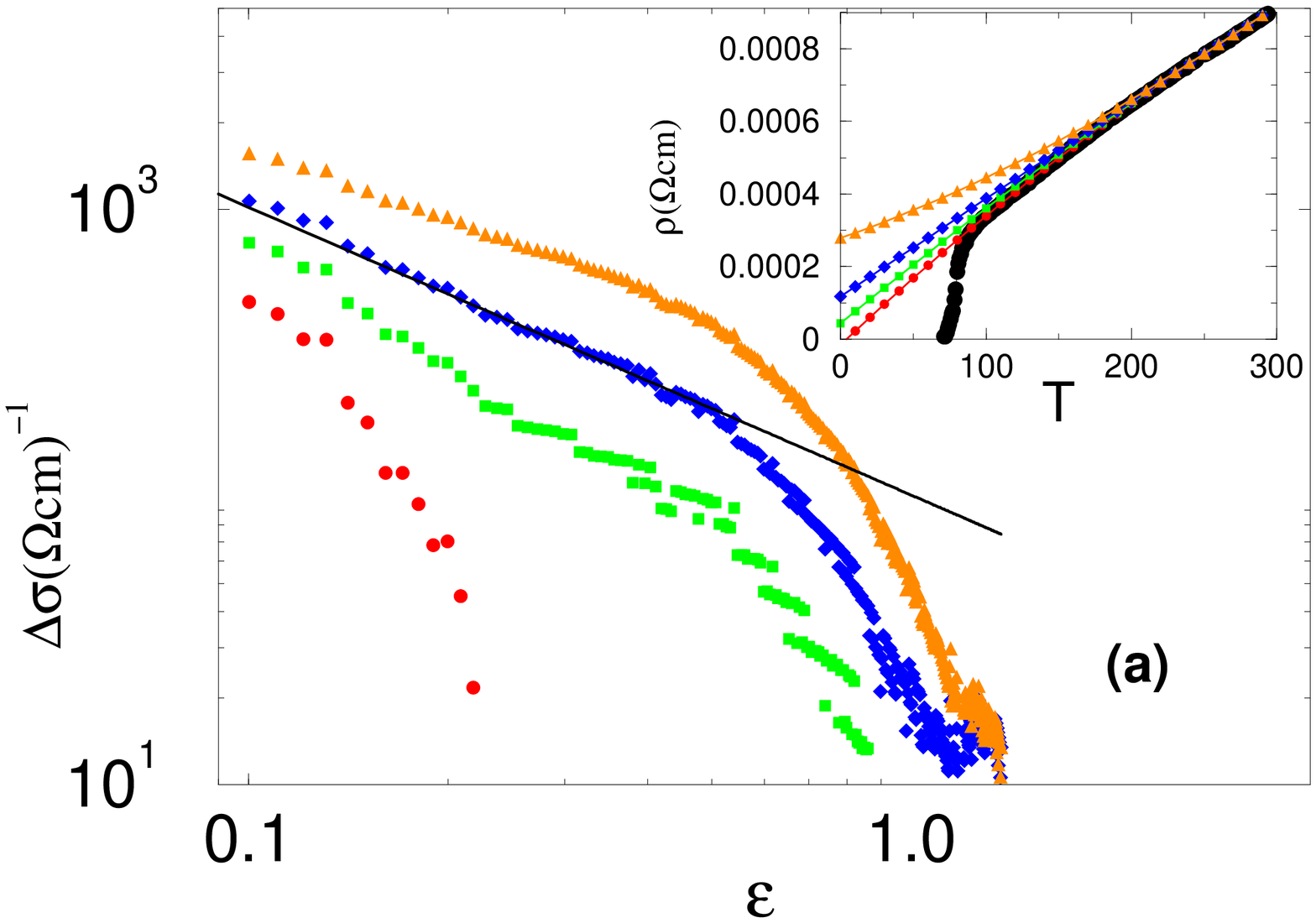}  
\includegraphics[angle=-90,scale=0.4]{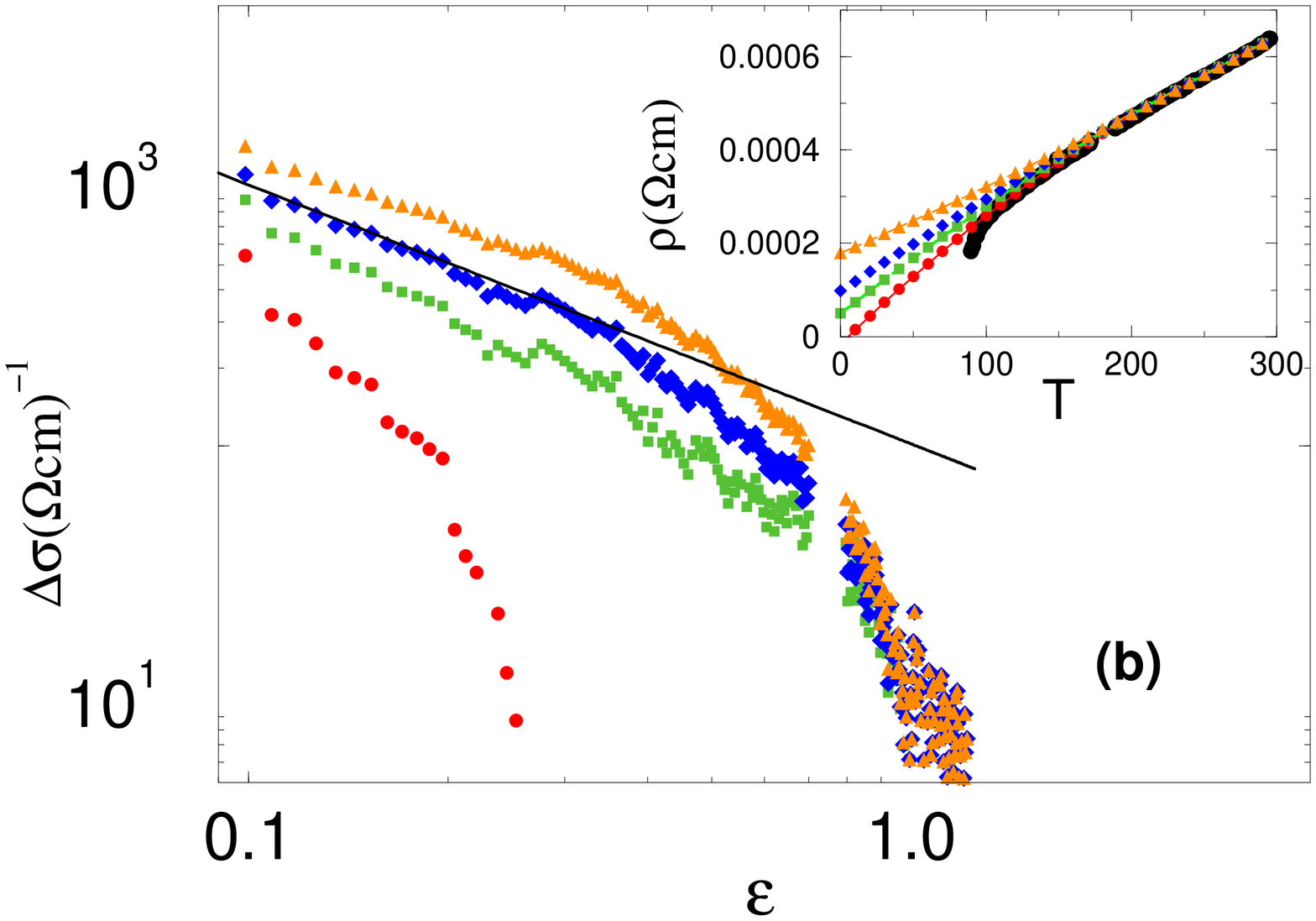}  
\caption{(Color online) 
Paraconductivity data, as obtained by taking 
the different normal-state resistivities. The resistivity data (black circles
in the insets) are from Ref. \onlinecite{WATANABE} for a BSCCO sample at 
doping (a) $x=0.217$, and (b) $x=0.22$.
Various hypotetical forms of the normal-state 
resistivity are reported in the insets.
All the curves coincide with a straight line above $T^*=250$ K in (a) and 
$T^*=220$ K in (b), 
while are quadratic below it. The color and the symbols of each set of 
paraconductivity points 
corresponds to the color and symbols of the normal-state resistivity curves 
of the insets. 
The black solid straight line is the universal 2D AL paraconductivity 
(no adjustable parameter).}
\label{Fig2}  
\end{figure}  
For each choice of the normal-state resistivity we determine the 
paraconductivity
$$
\Delta\sigma_{D=2}(T)= \frac{1}{\rho(T)}-\frac{1}{\rho_n(T)}
$$
obtaining the data of Fig. 1. The black solid straight line represents
the pure AL paraconductivity. Rather naturally, if one chooses the normal state
to follow closely the resistivity data [the red circles in the insets], there is 
little space for the paraconductivity contribution, which rapidly dies above 
$T_c$. Nevertheless, one can see that approaching $T_c$ the paraconductivity 
[the red circles in the main panels] merges with the (diverging) AL contribution. On 
the other hand one can choose the normal-state with an upward curvature [the 
orange triangles in the insets], which emphasizes the difference between the 
resistivity data and the (supposed) normal-state resistivity. In this case the 
paraconductivity must be large to bring the large normal-state resistivity 
down to the observed values. The orange triangles of the main panels 
represent this large 
contribution to the paraconductivity. In this case one sees that 
$\Delta\sigma_{D=2}(T)$ has the same slope as the pure AL paraconductivity,
but has a nearly constant positive offset and is rapidly suppressed around 
$\varepsilon \sim 0.5$, corresponding to $T\sim T^*$. This last effect simply 
arises from the ``perfect'' matching of the linear resistivity data with the 
assumed linear normal state resistivity for  $T>T^*$. In between the two 
limiting cases described above, there is the choice of normal state 
resistivities with small (or vanishing) curvature represented by the 
blue diamonds of the insets. Quite interestingly, one finds that the related 
paraconductivity closely follows the pure AL behavior, both for the slope and 
for the absolute (universal) value. This shows that the resistivity data, not 
only are compatible with a 2D AL behavior near $T_c$, but also this behavior 
extends up to $T^*$ provided a (nearly) linear  normal-state resistivity is 
assumed. Also in this case, as soon as the temperature reaches $T^*$, the 
paraconductivity rapidly drops.

This behavior is suggestive of the fact that below $T^*$ the resistivity
would be linear were it not for the presence of gaussian Cooper-pair 
fluctuations giving an AL contributions to the conductivity. These suppress the
resistivity below its linear behavior all over the $T<T^*$ region.

{\it --- Conclusions ---}  
In this paper we critically revisited the paraconductivity data in the 
cuprates addressing the two main issues: The existence and robustness of the 
AL paraconductivity, which in underdoped systems survives well above $T_c$, 
and the rapid suppression of paraconductivity above $T^*$. Regarding the 
second issue, we recast the problem of the cutoff in the pairing 
collective-mode fluctuations, showing that the rapid suppression of the pairing
fluctuations away from $T_c$ can arise from a rapid suppression of the
spectral weight of the pair fluctuations above a characteristic energy scale,
which directly involves $T^*$, $\Omega_0\equiv \gamma^{-1}\log(T^*/T_c)$.
As far as the second issue is concerned, we find the surprising result that,
assuming a (nearly) linear normal-state resistivity, the measured 2D 
paraconductivity in BSCCO closely follows the pure AL behavior. It seems to us 
that the coincidence (revealed at all dopings up to the optimal one) both for 
the power-law and the universal prefactors between the extracted 
paraconductivity and the AL behavior can hardly be casual. This suggests that 
the temperature dependence of the resistivity in BSCCO is given by
a normal-state linear contribution, which is decreased below $T^*$ by the 2D
AL paraconductivity. If, as it seems natural, this paraconductivity arises from
gaussian pair fluctuations, our analysis entails that preformed pairs, if any,
do not provide a separate additional conductivity channel.

We acknowledge interesting discussions with C. Castellani, 
C. Di Castro and M. Aprili. S. C. and M. G. acknowledge financial support from MIUR Cofin 
2003, Prot.  $2003020239\_006$, and from ESPCI/CNRS, which they also thank for 
the warm  hospitality.


\begin{thebibliography}{99}
\bibitem{notavidal}  This relevance of $T^*$ as a cutoff scale is 
at variance with the proposal of Ref. \onlinecite{vidal}, where
the energy cutoff of pair fluctuations was related to $\xi_0$.
\end{thebibliography}
\end{document}